\documentclass[authoryear,preprint,11pt]{elsarticle}

\usepackage[left=1in,top=1in,right=1in,bottom=1in]{geometry}

\usepackage{graphicx}

\usepackage{amssymb}

\usepackage[displaymath]{lineno}

\usepackage{bm}

\usepackage{subeqn}

\newcommand{\beq}{\begin{equation}}
\newcommand{\eeq}{\end{equation}}
\newcommand{\bea}{\begin{eqnarray}}
\newcommand{\eea}{\end{eqnarray}}
\newcommand{\bes}{\begin{subeqnarray}}
\newcommand{\ees}{\end{subeqnarray}}
\newcommand{\dfrac}[2]{\displaystyle\frac{#1}{#2}}

\newcommand{\mten}[1]{\mathsf{#1}}
\newcommand{\mvec}[1]{\boldsymbol{#1}}

\newcommand{\mrm}[1]{\mathrm{#1}}
\newcommand{\del}{\nabla}
\newcommand{\divr}{\nabla \cdot}
\newcommand{\curl}{\nabla \times}
\newcommand{\epsi}{\epsilon}
\newcommand{\vepsi}{\varepsilon}
\newcommand{\hhat}{\hat{\mvec{h}}}

\newcommand{\phihat}{\hat{\mvec{\phi}}}
\newcommand{\Zhat}{\hat{\mvec{Z}}}
\newcommand{\Rhat}{\hat{\mvec{R}}}
\newcommand{\perphat}{\hat{\mvec{\perp}}}
\newcommand{\parahat}{\hat{\mvec{\parallel}}}
\newcommand{\para}{\parallel}
\newcommand{\rhat}{\hat{\mvec{r}}}
\newcommand{\zhat}{\hat{\mvec{z}}}
\newcommand{\thetahat}{\hat{\mvec{\theta}}}

\newcommand{\wt}[1]{\widetilde{#1}}

\newcommand{\oover}[1]{\dfrac{1}{#1}}
\newcommand{\dext}{\mathrm{d}}

\newcommand{\partlxy}[2]{\partial {#1} / \partial {#2}}

\newcommand{\Cth}{\cos \theta}
\newcommand{\Sth}{\sin \theta}
\newcommand{\CCth}{\cos 2\theta}
\newcommand{\SSth}{\sin 2\theta}
\newcommand{\Epol}{E_\theta}
\newcommand{\Etor}{E_\phi}
\newcommand{\Bpol}{B_\theta}
\newcommand{\Btor}{B_\phi}

\newcommand{\Vpolj}{V_{\theta\, j}}
\newcommand{\Vtorj}{V_{\phi\, j}}

\newcommand{\ddr}[1]{\dfrac{\partial\, {#1}}{\partial r}}
\newcommand{\ddp}[1]{\dfrac{\partial\, {#1}}{\partial \theta}}
\newcommand{\ddpor}[1]{\dfrac{\partial\, {#1}}{r \partial \theta}}

\newcommand{\ddZ}[1]{\dfrac{\partial\, {#1}}{\partial Z}}
\newcommand{\ddR}[1]{\dfrac{\partial\, {#1}}{\partial R}}

\journal{Mechanics Research Communications}

\begin{document}

\begin{frontmatter}



\title{Critical evaluation of the neoclassical model for the equilibrium electrostatic field in a tokamak}


\author{Robert W. Johnson}
\ead{robjohnson@alphawaveresearch.com}
\address{Alphawave Research, 29 Stanebrook Ct., Atlanta, GA 30238, USA}

\begin{abstract}
The neoclassical prescription to use an equation of motion to determine the electrostatic field within a tokamak plasma is fraught with difficulties.  Herein we examine two popular expressions for the equilibrium electrostatic field so determined and show that one fails to withstand a formal scrutiny thereof while the other fails to respect the vector nature of the diamagnetic current.  Reconsideration of the justification for the presence of the equilibrium electrostatic field indicates that no field is needed for a neutral plasma when considering the net bound current defined as the curl of the magnetization.  With any shift in the toroidal magnetic flux distribution, a dynamic electric field is generated with both radial and poloidal components, providing an alternate explanation for any measurements thereof.
\end{abstract}


\end{frontmatter}

\linenumbers


\section{Introduction} \label{sec:intro}
The magnetic confinement of burning hydrogenic plasma is of primary concern to those interested in the production of power from nuclear fusion.  The most popular approach is based on the tokamak design~\citep{diiid-2002}, in which a current is driven through an ionized gas within a toroidal confinement vessel surrounding a central coil by transformer action.  Some geometric nomenclature and useful coordinate relations are given in the Appendix.  The vacuum vessel is encircled by poloidal coils which produce the toroidal magnetic field, and the toroidal plasma current produces the poloidal magnetic field.  The mathematical description of the device is generally based on that of plasma physics, which in principle is a marriage of kinetic mechanics and electrodynamics.  In practice, however, the standard description of the plasma phase of matter neglects one very important equation, thus introducing mathematical inconsistency to the model.

Different authors use the term ``neoclassical'' to mean different things; most often it is applied to theories of particle transport in toroidal geometry, as opposed to ``classical'' transport in an axial geometry.  We prefer to use the term to include all theories of ionized media which are non-classical by construction, \textit{ie} not based on the potential formulation of electrodynamics, whereby the so-called quasineutral approximation (which is not an approximation but rather an assumption on the mathematical form of the theory) requires the use of an equation of motion to determine the electric field rather than Gauss's law or Poisson's equation~\citep{chen-84,dendybook-93,tokamaks-2004}.  Plasma theorists have been neglecting Gauss's law since the advent of the topic's self-identification as a separate branch of physics~\citep{roseclark,brag-1965}, which may explain why the promise of fusion power has been only fifty years away for over fifty years.

Herein we examine two popular expressions for the electrostatic field determined by neoclassical models of plasma equilibrium, the first established from electron momentum conservation and the second from the Pfirsch-Schl\"{u}ter current, and show that one fails to withstand a formal scrutiny thereof and the other fails to respect the vector nature of the diamagnetic current.   Please forgive the thoroughness of the presentation, which addresses the arguments made by various colleagues in defense of the neoclassical approach.
At issue is the validity of the quasineutral approximation, which allows for a divergenceful electric field in the absence of a non-vanishing space charge density $\rho_e \equiv \sum_s n_s e_s$, formally expressed as $\divr \mvec{E} \neq 0$ for $\rho_e = 0$, and requires the determination of the electrostatic field from an equation of motion.  However, the quasineutral approximation does not respect the mathematics of electrodynamic field theory, \beq
\divr \mvec{E} = \rho_e / \vepsi_0 \;.
\eeq  From a particle physicist's perspective~\citep{davis70,ryder-qft,halzenmartin,ramond-FTM90,mandlshaw}, the gauge invariant Maxwell field tensor $F^{\mu \nu} \equiv \partial^\mu A^\nu - \partial^\nu A^\mu$ is known to carry only 3 scalar degrees of freedom in media, not 3 for each of the electric and magnetic fields, embodied by the four-potential $A^\mu \equiv (\Phi/\mrm{c},\mvec{A})$ subject to the gauge condition and coupled to sources given by the conserved four-current $J^\mu \equiv (\mrm{c} \rho_e,\mvec{J})$ through the inhomogeneous Maxwell equations $\partial_\nu F^{\mu \nu} = \mu_0 J^\mu$, while the homogeneous equations are recognized as the Bianchi identity for electromagnetism given by the field equation for the dual tensor $\partial_\nu \widetilde{F}^{\mu \nu} = 0$ and are satisfied identically when written in terms of the electromagnetic potential hence do not determine any degrees of freedom, thus the electrostatic field is determined by the space charge density $\rho_e$ and not by an equation of motion.  The equations by Maxwell may be expressed succinctly using intrinsic, geometric notation as $\dext\, ^* \dext\, A = J$ in terms of the exterior derivative $\dext$, the Hodge dual $^*$, the connection 1-form $A$, and the current 3-form $J$, as given in many standard quantum field theory texts, such as~\citet{ryder-qft}, or more esoteric monographs, such as~\citet{davis70}.  What that expression states is that Gauss's law may not be isolated from the remainder of the source bearing Maxwell field equations, $\curl \mvec{B} - \mu_0 \vepsi_0 \partial \mvec{E} / \partial t = \mu_0 \mvec{J}$, as they are but one unit of truth.

Note that we are not criticizing the kinetic approach to plasma calculations, which in its original inception as the Vlasov-Maxwell system~\citep{vlasov-68} fully respects the microscopic electrodynamic field theory, but rather the neoclassical (non-classical) fluid model based on the quasineutral approximation, which does not respect the macroscopic electrodynamic field theory.  While various approximations are made in the drift kinetic equations~\citep{hinton-836,lin-2975}, most practical numerical evaluations address Poisson's equation directly~\citep{belli-095010}.  With this article, we examine in detail the mathematical difficulties one encounters when following the neoclassical prescription.  Similar discussions on the existence of the whistler oscilliton in geophysical plasmas~\citep{npg-12-425-2005,npg-14-543-2007,npg-14-49-2007,npg-14-545-2007} and of the helicon wave in propulsion devices~\citep{rwj-pop01,walker-054702} are noted.  Criticism of the quasineutral approach in a cosmological context has also recently appeared~\citep{teodoro-383,diver-4632}, and it is time for the fusion community to address these difficulties head on.

We close by considering an alternate model for the production of electric fields within toroidal confinement devices based upon the physics of macroscopic electromagnetism.  Relaxing the requirement of equilibrium slightly indicates that changes to the kinetic pressure distribution (for constant net pressure) produce changes to the plasma magnetization which induce an electric field by Faraday's law.  Regions of opposite variation experience a mutually repulsive force, which may be a source of plasma instability not previously recognized in the literature.

\section{Electrostatic field from the species equations of motion} \label{sec:field1}
The first equilibrium $\partial / \partial t \rightarrow 0$ electrostatic field under consideration is one commonly used in the analysis of tokamak experiments~\citep{solomonetal-pop-2006,frc-pop-2006,stacey-cpp06,stacey012503}, determined by integration of the electron poloidal and ion radial equations of motion in the large aspect ratio, concentric circular flux surface approximation.  The use of concentric circular flux surfaces with a toroidal integrating measure is pursued herein to remain consistent with the model as presented in the literature, as are the expansions in $\epsi \equiv r/R_0$ of the electron density $n_e = n_e^0(r) [ 1 + n_e^c(r) \Cth + n_e^s(r) \Sth ]$ and electrostatic potential.  \citet{stacey012503} state that ``the electron momentum balance can be solved for $\wt{\Phi}^{c,s} \equiv \Phi^{c,s}/\epsi = n_e^{c,s}/\epsi(e \Phi^0 / T_e)$, which represents the poloidal asymmetry in the electrostatic potential.''  Let us examine that statement in detail.

\subsection{Reduction to homogeneous form}\label{ssec:homoeqn}
This neoclassical model considers the equilibrium poloidal equation of motion for arbitrary species $s$ to be \bes 
0 & = & \left[n_s m_s \left(\mvec{V}_s \cdot \del \right)\mvec{V}_s + \divr \mten{\Pi}_s \right] \cdot \thetahat \nonumber \\
 & & + \partial p_s / r \partial \theta - F_{s\, \theta} + n_s e_s \left(V_{s\, r} \Btor - \Epol\right) \;,
\ees where $p_s = n_s T_s$ for $T_s \leftarrow k_B T_s$ and $\mvec{F}_s$ is the friction term, and takes the poloidal component of the electrostatic field $\Epol \equiv - \partial \Phi(r,\theta) / r \partial \theta$ on a flux surface at $r$ in Coulomb gauge as \bes
\Epol & = & -  \ddpor{}\Phi^0(r) \left[ 1+\Phi^c(r)\Cth+\Phi^s(r)\Sth \right] \\
 & = & -  \dfrac{\Phi^0(r)}{r}\left[\Phi^s(r)\Cth-\Phi^c(r)\Sth\right] \;,
\ees where $\Phi$ is the electrostatic potential, indicating an expansion around $\Phi^0(r) \equiv -\int_a^r dr E_r^0 \neq 0$ for a last closed flux surface at $r=a$, where the radial electrostatic field is calculated from an ion equation of motion~\citep{solomonetal-pop-2006}.  The resulting evaluation of the flux surface unity, cosine, and sine moments of the electron poloidal equation of motion with $\partial T_e / \partial \theta = 0$ (where other terms are assumed negligible at equilibrium), \beq \label{eqn:epol}
T_e \partial n_e / \partial \theta = -e n_e r \Epol \;,
\eeq defined by the expressions \beq
\langle A \rangle_{\{U,C,S\}}\equiv \dfrac{1}{2 \pi} \oint d\theta \{1,\Cth,\Sth\}(1+\epsi\Cth) A \;,
\eeq yields three equations which have only trivial solution.  Specifically, we have the system of equations \bes \label{eqn:U}
 U: & \; \epsi n_e^s T_e &\! =\, e \Phi^0 (\epsi \Phi^s + n_e^c \Phi^s - n_e^s \Phi^c)  \;, \\ \label{eqn:C}
 C: & \; n_e^s T_e &\! =\, e \Phi^0 (4\Phi^s + 3 \epsi n_e^c \Phi^s - \epsi n_e^s \Phi^c)/4  \;, \\ \label{eqn:S}
 S: & \; n_e^c T_e &\! =\, e \Phi^0 (4\Phi^c + \epsi n_e^c \Phi^c - \epsi n_e^s \Phi^s)/4  \;, 
\ees valid $\forall \; \epsi,n_e^c,n_e^s$, and $T_e$.  The terms with factors of $\epsi$ above are strictly due to the toroidal geometry and would disappear for a cylindrical plasma column $R_0 \rightarrow \infty$; one cannot address the extension to toroidal geometry of other aspects of the model~\citep{stacandsig-1985} without addressing the extension here.  Solution in pairs given finite (fixed) $\Phi^0$ yields inconsistent values of $\Phi^{c,s}$ and an overdetermined system, which therefor has no solution, thus the poloidal electrostatic field in this neoclassical model, which fails to consider the $O(\epsi)$ terms within the cosine and sine moment equations, is unphysical.  Failing to include the $O(\epsi)$ terms indicates expressions applicable only on the magnetic axis $r=0$, where $\epsi = 0$ for $R_0 \neq \infty$, yet this neoclassical model is commonly used to address the physics near the edge of the confinement region~\citep{stacey012503}.  This system may be put into linear, homogeneous form $\mten{A} \mvec{x} = 0$ by dividing through by $\Phi^0$, \beq \label{eqn:linepol}
\left[\begin{array}{ccc} -\epsi n_e^s T_e / e & - n_e^s  & \epsi + n_e^c \\ -4 n_e^s T_e / e & - \epsi n_e^s & 4 + 3 \epsi n_e^c \\ -4 n_e^c T_e / e & 4 + \epsi n_e^c & - \epsi n_e^s  \end{array} \right] 
\left[\begin{array}{c} 1/\Phi^0 \\ \Phi^c \\ \Phi^s \end{array}\right] = \left[\begin{array}{c} 0 \\ 0 \\ 0\end{array}\right] \;,
\eeq  thus its {\it only} exact solution for $\epsi \neq 0$ is trivial, \beq
(1/\Phi^0,\Phi^c,\Phi^s) \equiv (0,0,0) \;,
\eeq which we interpret to mean exactly what it says, that it is solved when $\Phi^0(r) = \pm \infty$, displaying its unphysical definition when determined from the electron poloidal equation of motion.  The matrix $\mten{A}$ has rank 3, thus the solution has 0 free parameters~\citep{schneider-68}.  For a cylindrical column $\epsi \rightarrow 0$, one recovers a matrix of rank 2 and the solution $\Phi^{c,s} = n_e^{c,s} (T_e / e \Phi^0)$ with $\Phi^0$ a free parameter.  If one assumes that the smallness of the coefficients may be represented by $\epsi$, {\it eg} $n_e^{c,s} \equiv \epsi \wt{n}_e^{c,s}$, then the offending terms near the edge where $\epsi$ approaches 1/2 represent up to a 19\% correction to the leading order equations.  As non-vanishing $\Phi^{c,s}$ are an integral part of the development of this neoclassical model and appear in its remaining equations without a prefactor of $\Phi^0$ through substitution, the validity of its conclusions is in jeopardy.  Note that a putative non-vanishing radial electrostatic field without poloidal variation demands the existence of {\it no} poloidal electrostatic field, else the poloidal variation to the potential ruins the poloidal symmetry of the radial field; if that radial field is determined from a radial equation of motion then the associated poloidal field is determined by the poloidal dependence of that equation.

One might think to alleviate the difficulty by invoking the logarithmic derivative, writing the electron poloidal equation of motion as $\partial (\ln n_e - e \Phi / T_e)/\partial \theta = 0$, with solution $C(r) = n_e(r,\theta) \exp [- e \Phi(r,\theta) / T_e(r)]$.  Expanding the exponential gives, for $-\infty < e \Phi / T_e < \infty$, \beq \label{eqn:Ceqn1}
C(r) = n_e \sum_{k=0}^\infty \frac{(- e \Phi / T_e)^k}{k!}
\eeq and to each order in $e \Phi / T_e$, taking the flux surface moments yields three equations to solve.  
The difficulty with $\Phi^0$ encountered above has simply been shifted to the ``undetermined'' function $C(r)$, which is perfectly determinable in principle from the system of equations.  
As the electron density $n_e^0$ divides out of Equation~(\ref{eqn:epol}), its poloidal variations $n_e^{c,s}$ are determined by continuity $\langle \partial n_e / \partial t + \divr n_e \mvec{V}_e - \dot{n}_e \rangle_{C,S} = 0$ for particle source rate $\dot{n}$, and the thermal energy $T_e$ is determined by the heat equation and given as input from experimental measurement for the analysis; the remaining degree of freedom $\Phi^0$ must be determined by $C(r)$.  
Concern over the expansion in no way detracts from the observation that the physics of the situation is embodied by the algebraic Equation~(\ref{eqn:linepol}) which contains no exponential factor thus represents a more exact method of solution.  

\subsection{Two species, one field}\label{ssec:onefield}
Next, we consider the consistency of using two different species' equations of motion to determine the species independent electrostatic potential, taking the radial electrostatic field from the equation of motion for arbitrary ion species $j$ as \beq
E_r  = \oover{n_j e_j}\ddr{p_j} + \Vtorj \Bpol - \Vpolj \Btor \;,
\eeq and dropping the convective term as is standard practice in the field.  The usual evaluation of that expression from experimental measurements neglects any poloidal dependence; however, when the intrinsic variation of quantities induced by the geometry is considered~\citep{mingagrees}, given in the large aspect ratio $\epsi \ll 1$, concentric circular $R_r \equiv R_0$, flux surface approximation $\mvec{B} = (0,\Bpol,\Btor)$ by \beq
\begin{array}{ll}
\mvec{B}=\mvec{B}^0/(1+\epsi\Cth) \;, & n_j = n_j^0 \;, \\
\Vpolj=\Vpolj^0/(1+\epsi\Cth) \;, & \Vtorj=\Vtorj^0(1+\epsi\Cth) \;, \end{array}
\eeq where $A^0$ is the average of the values of $A$ on the vertical midplane, a geometric dependence is introduced.  Using these values ($r$ dependence implied), we find \beq
E_r(r,\theta) = \dfrac{p'^{\,0}_j}{n^0_j e_j} + \Vtorj^0 \Bpol^0 - \Vpolj^0 \Btor^0 /(1+\epsi\Cth)^2\;,
\eeq for species pressure gradient $p'_j \equiv \partial n_j T_j / \partial r$, thus $E_r(\theta) \not\equiv E_r^0$.  Expanding the denominator reveals a power series in $\epsi\Cth$, \bes
E_r(r,\theta) & = & \dfrac{p'^{\,0}_j}{n^0_j e_j} + \Vtorj^0 \Bpol^0 - \Vpolj^0 \Btor^0 \sum_{k=0}^\infty [-(k+1)]^k \epsi^k \cos^k \theta \nonumber \\ \\ 
& \approx & \dfrac{p'^{\,0}_j}{n^0_j e_j} + \Vtorj^0 \Bpol^0 - \Vpolj^0 \Btor^0 + 2 \epsi \Vpolj^0 \Btor^0 \Cth \\
& \approx & E_r^0(r) + E_r^1(r) \Cth \;,
\ees and as integration with respect to $r$ does not affect the $\theta$ dependency, the potential associated with the radial electrostatic field relative to its central value, $\Phi_{E_r}(r,\theta) \equiv - \int_0^r dr E_r(r,\theta)$, may be written as a cosine series, \beq \label{eqn:phier}
\Phi_{E_r}(r,\theta) \approx \Phi_{E_r}^0 + \Phi_{E_r}^1 \Cth = \Phi_{E_r}^0 \left(1 + \Phi_{E_r}^c \Cth \right) \;,
\eeq where $\Phi_{E_r}^c = \int_0^r dr E_r^1 / \int_0^r dr E_r^0 \neq 0$ in general, noting that the potential on the last closed flux surface at $r=a$ is {not} single valued.  Returning now to the electrostatic potential appearing in the poloidal equation of motion, expressed to leading order as \beq \label{eqn:phiepol}
\Phi_{\Epol}(r,\theta) = \Phi_{\Epol}^0 (r) [1 + \Phi_{\Epol}^c (r) \Cth  + \Phi_{\Epol}^s (r) \Sth] \;,
\eeq where $\Phi_{\Epol}^0(r) \equiv -\int_a^r dr E_r^0 \neq 0$ is the potential relative to that of the last closed flux surface, with solution $\Phi^{c,s}_{\Epol} = n_e^{c,s} (T_e / e \Phi)$.  No loss of generality ensues if one redefines the potential relative to its central value.  With vanishing extrinsic poloidal dependence to the electron density, $n_e^{c,s} \rightarrow 0$ as above, the potential retains no explicit poloidal dependence, $\Phi_{\Epol}(r,\theta) \rightarrow \Phi_{\Epol}^0 (r)$.  Thus, we conclude that the electrostatic potentials associated with the radial and poloidal electrostatic fields evaluated from the ion and electron equations of motion by this neoclassical model are inconsistent, $\Phi_{E_r} \neq \Phi_{\Epol}$, as Equation~(\ref{eqn:phier}) does not equal Equation~(\ref{eqn:phiepol}) in the case of vanishing density asymmetries $n_{e,j}^{c,s} = 0$.

\section{Electrostatic field from the Ohm's law equation} \label{sec:field2}
Examining the expression of another leading contender for the equilibrium electrostatic field~\citep{tokamaks-2004} evaluated from the Ohm's law equation and the Pfirsch-Schl\"{u}ter current, one may put its poloidal component \bes \label{eqn:psepol}
\Epol & = & \dfrac{\langle\Etor\Btor/\Bpol\rangle B^2}{\langle B^2/\Bpol\rangle \Bpol} - \dfrac{\Etor\Btor}{\Bpol} \nonumber \\
 & & + R\Btor p' \eta_\para \left(\dfrac{\langle 1/\Bpol\rangle B^2}{\langle B^2/\Bpol\rangle \Bpol} -\dfrac{1}{\Bpol}\right) \;,
\ees into the form \beq \label{eqn:neoepol}
\Epol = \Epol^c \left[ 2\epsi\Cth - (\epsi^2/2) \CCth \right] \;,
\eeq when the Shafranov shift is neglected, as in the concentric circular flux surface approximation of above, upon application of Stokes' theorem to Faraday's law, {\it ie} by requiring $\oint d\theta \Epol=0$.  Note that this neoclassical model for the poloidal electrostatic field differs distinctly from that of the previous section in detailed functional form.  Inserting Equation~(\ref{eqn:neoepol}) into Equation~(\ref{eqn:epol}) and taking the flux surface Fourier moments yields three equations which have a nontrivial solution only when expanded to order $O(\epsi^3)$, given by \beq \label{eqn:neoepolsoln}
\left[\begin{array}{c} n_e^c \\ n_e^s \\ \Epol^c\end{array}\right] = \left[\begin{array}{c} \epsi^3/(6\epsi^2-8) \\ \pm\epsi\sqrt{3\epsi^4-168\epsi^2+192}/(18\epsi^2-24) \\ \pm4(T_e/e R_0)/\epsi\sqrt{3\epsi^4-168\epsi^2+192}\end{array}\right]\;,
\eeq thus the presence of a poloidal electrostatic field of that form should be accompanied by a potentially measurable shift in the electron density profile.  

Note that the derivation immediately preceding is slightly inconsistent, as the associated electrostatic potential $\Phi_\mrm{axi} = \Phi_0 [ 2 \epsi \Sth - ( \epsi^2 / 4 ) \SSth ]$ is of the correct harmonic form~\citep{flanigan,book-nrl-77,simmons-91,BLT-92} for axial geometry, as is easily verified in $(Z=-r\Sth,R=R_0+r\Cth,z)$ coordinates via application of the axial Laplacian $\del^2_\mrm{axi}\equiv\partial^2/\partial Z^2+\partial^2/\partial R^2$ to $\Phi_\mrm{axi}=\Phi_0 Z [(R-R_0)/2R_0^2-2/R_0]$, yet the flux surface average is done in toroidal geometry.  The harmonic potential for the toroidal Laplacian $\del^2_\mrm{tor}\equiv\del^2_\mrm{axi}+\partial/R\partial R$ is written $\Phi_\mrm{tor} = \Phi_\mrm{axi}(R\rightarrow\ln R)$, from which $E_Z=-\Phi_0[(\ln R-R_0)/2R_0^2-2/R_0]$ and $E_R=-\Phi_0 Z/2RR_0^2$, noting that the introduction of the logarithm breaks the usually obvious relation between the symbol for the magnitude of a quantity and the units associated with that quantity---carefully pulling the units beside the leading coefficients of expressions ensures that they are respected.  Note that this $\Phi_0$ is {\it not} the $\Phi^0$ of the preceding section but is a unit bearing constant which sets the scale.  
From these, we determine the poloidal field $\Epol \equiv -\partial \Phi / r \partial \theta$ to be \bes
\Epol & = & \left(\frac{-\Phi_0}{r}\right)\left(\ddp{Z}\ddZ{}+\ddp{R}\ddR{}\right)\frac{\Phi}{\Phi_0} \;, \\
 & = & \Epol^c\left[(R-R_0)\frac{E_Z}{\Phi_0}-Z\frac{E_R}{\Phi_0}\right]\;, \ees
where we identify $E_{r,\theta}^c\equiv-\Phi_0/r$, and the corresponding radial field $E_r \equiv -\partial \Phi / \partial r$ is \bes
E_r & = & \left(\frac{-\Phi_0}{r}\right)r\left(\ddr{Z}\ddZ{}+\ddr{R}\ddR{}\right)\frac{\Phi}{\Phi_0} \;, \\
 & = & E_r^c\left[-Z\frac{E_Z}{\Phi_0}-(R-R_0)\frac{E_R}{\Phi_0}\right]\;. \ees
In $(r,\theta,\phi)$ coordinates, we have $\Phi_\mrm{tor}=\Phi_0 r \Sth [-\ln(R_0+r \Cth)+5 R_0]/2R_0^2$.  
As this is an electrostatic field within a neutral medium ({\it ie} one for which the net charge on a differential volume element vanishes) at equilibrium, Maxwell's equations $\curl\mvec{E}=0$ and $\divr\mvec{E}=0$ are satisfied within the bulk region.

The Pfirsch-Schl\"{u}ter current is supposed to flow along the electrostatic field of this section to cancel the charge accumulation arising from the pressure gradient driven diamagnetic current $\mvec{J}_{\del p} = - \del p \times \mvec{B}/B^2$ in toroidal geometry~\citep{tokamaks-2004}.  For the harmonic potential, these charges accumulate 
on the boundary of the region under consideration, which in this case is the $R/R_0$ weighted circle representing our outermost flux surface at normalized minor radius $r/a=1$ upon collapse of the toroidal dimension,  
writing $ -\partial \rho_{\del p} / \partial t = \del_\mrm{tor} \cdot \mvec{J}_{\del p} \neq 0$.  The motivation for this accumulation is \bes
\del_\mrm{tor} \cdot  \mvec{J}_{\del p} & = & \divr \left(\dfrac{\mvec{B}\times\del p}{B^2}\right) \;, \\
 & = & \dfrac{\del p \cdot (\curl\mvec{B}) - \mvec{B} \cdot(\curl\del p)}{B^2} \nonumber \\
 & & + (\mvec{B}\times\del p) \cdot \del \oover{B^2}  \;,
\ees where the final term is nonzero due to the poloidal dependence of $B$.  However, $\mvec{J}_{\del p}$ is but one component of the total diamagnetic current, properly defined as the curl of the magnetization $\mvec{J}_\mrm{dia} \equiv \curl \mvec{M}$ where $\mvec{M} \equiv - (p/B^2)\mvec{B}$, which includes the effects of the pressure gradient driven current as well as the curvature and $\del B$ drift currents~\citep{hazeltine-04} and remains divergence-free regardless of the geometry $\divr \mvec{J}_\mrm{dia} \equiv 0$, thus there is no space charge accumulation and no motivation for a canceling current.  (The common procedure of adding the particle drift currents to the fluid diamagnetic current we feel represents an over-counting of the underlying phenomenon, as all currents are the curl of either an $\mvec{H}$ or an $\mvec{M}$.)  The error here lies in not fully distinguishing the free and bound charges and currents as they appear in the Heaviside notation of the Maxwell equations and in trying to model a fully ionized medium as both a conductor and dielectric at zero frequency.

\section{Dynamic electric field}
Relaxing our requirement of equilibrium slightly, we consider now the effect of a shift in the magnetic flux density when the total magnetic flux remains constant.  This effect may possibly account for the direct experimental observations of an electric field within the device~\citep{cortes-1596,holcomb-10E506}.  As Gauss's law is inviolate, any electric fields within the neutral, conducting medium of a tokamak plasma necessarily are driven by changes in the magnetic flux density at some location in space, giving us \beq \label{eqn:E}
\divr \mvec{E} = 0\;,\;\curl \mvec{E} = -\partlxy{\mvec{B}}{t} \;.
\eeq  These changes may result for fixed $H_\phi$ by a shift in the pressure distribution leading to $\partial M_\phi / \partial t \neq 0$, as $\mvec{B}/\mu_0 = (H - M) \hhat$.  Noting the similarity to the laws of Thomson and Ampere allows one to define an electric vector potential $\mvec{E} = \curl \mvec{F}$ in the analogue of Coulomb gauge $\divr \mvec{F} = 0$.  Then $\nabla^2 \mvec{F} = \partlxy{\mvec{B}}{t}$ has the solution~\citep{Laslett-1987bq,jackson-third,deolive-1175,rwj-epjd01} for a circular loop source at $(Z_0,R_0)$ given by \beq
F_\phi / F_\phi^0 = \frac{4\sqrt{a+b}}{b} \left[ \left( \frac{a}{a+b} \right) K(k) - E(k) \right] \;,
\eeq expressed in terms of the complete elliptic integrals~\citep{abramowitz-stegun} with parameter $k^2 = 2 b / (a+b)$, where $a=(Z-Z_0)^2 + R^2 + R_0^2$ and $b = 2 R R_0$ and its magnitude \beq
F_\phi^0 = - (\partlxy{B_\phi}{t}) \Delta^2 R_0 / 4 \pi
\eeq is supposed constant over a differential area $\Delta^2$.

\begin{figure*}%
\includegraphics[width=.6\textwidth]{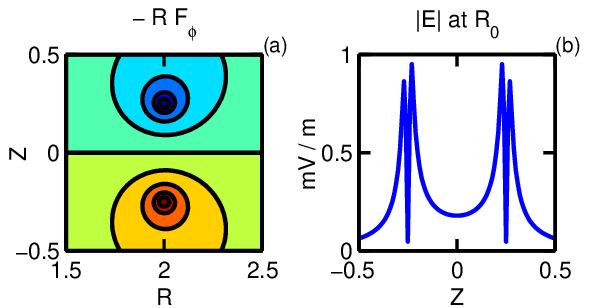}
\caption{Electric field generated from changing magnetization for parameters given in the text. Contours (a) of the flux function follow the electric field. Magnitude (b) of the electric field along the vertical midplane at $R_0 = 2$.}
\label{fig:C}
\end{figure*}

On a uniform grid $(Z,R)$ in meters with spacing $\Delta = .01$~m, we wish to evaluate $F_\phi$ for sources with opposite polarity around $(\pm Z_0,R_0)$ for $Z_0 = 0.25$ aligned to the vertical midplane at $R_0 = 2$.  Expressing the plasma magnetization~\citep{rwj-jpp03,rwj-mrc01} as $M/H = (1-\sqrt{1-4 p / \mu_0 H^2})/2$ for $\hhat = \phihat \cos \zeta + \thetahat \sin \zeta$ and supposing the pitch angle $\zeta = \epsi \pi / 2$ for $\epsi = Z_0 / R_0$ and $\partlxy{\mvec{H}}{t} = 0$ allows one to write \beq
\partlxy{B_\phi}{t} = - (\cos \zeta) (H^2 - 4 p / \mu_0)^{-1/2} \partlxy{p}{t} \;.
\eeq  Using some typical parameters $n_0 = 10^{19}\,\mrm{m}^{-3}$, $T_e = 3$~keV, $T_i = 9$~keV, $B_\phi = 2$~T, $B_\theta = 0.2$~T, and supposing the rate of change to the pressure is 1\% per millisecond, lets one give a numerical estimate to $F_\phi^0$ hence the magnitude of the electric field.  From Gauss's law one defines the flux funtion $\chi$ such that \beq
\maltese \chi \equiv \phihat \times \del \chi = R \mvec{E} = R \curl \mvec{F} \;,
\eeq whence $\chi = - R F_\phi$ upto an unphysical constant, where $\maltese$ is the contour operator orthogonal to $\del$ and $\phihat$.  The results of this evaluation are shown in Figure~\ref{fig:C}, where one sees that an electric field with magnitude approaching 1~mV/m for these parameters may arise on the vertical midplane.  Pursuing the analogy between $\mvec{E} \sim \mvec{B}$ and $\mvec{J} \sim -\partlxy{\mvec{M}}{t}$, a force of repulsion \beq
\mvec{F}_\pm = - \mu_0 \vepsi_0 (\partlxy{\mvec{M}_\pm}{t}) \times \mvec{E}_\mp
\eeq should appear between line sources of opposite polarity by action of the macroscopic Lorentz force~\citep{mansuripur-1619,mansuripur-1608}, where the subscript indicates the source location, thus a perturbation to the pressure does not necessarily coalesce.

\section{Conclusions and outlook} \label{sec:conc}
From the preceding analysis, we find that the use of an equation of motion to determine the equilibrium electrostatic field rather than Poisson's equation leads to formal inconsistencies in the model.  Part of the problem lies in treating the poloidal and radial components of the field separately, when there is only one electrostatic potential from which both components may be determined.  The remainder lies in the neglect of Gauss's law popularly established within the plasma physics community~\citep{goldruth95,kivel95,spsbds03,mbk04,dinkbook-05}, which relegates the defining relation for the electrostatic field to a position of subsidiary moment.  Any model which treats the plasma as a neutral, conducting fluid, where neutrality is understood to hold down to some scale smaller than the differential volume element used to define the continuum quantities, needs to respect {\it all} of Maxwell's laws, which are manifestly Lorentz covariant.  Taking a field theoretic perspective implies that the electrostatic field within a neutral medium must either vanish or result from sources and be supported by dielectric polarization.  

In conclusion, we have found that neither neoclassical model for the equilibrium electrostatic field within a tokamak respects the mathematics of electrodynamic field theory.  An alternative model relating changes in pressure to changes in magnetization may explain observations of an electric field during experiments.  The potential formulation is becoming widely recognized as more fundamental even at the classical level, while the macroscopic field formulation can be just as useful for the description of nonlinear media as for linear ones.  The thrust of our argument is simply that, if there is an electric field within a tokamak, then it must follow the same rules as every other electric field found in nature.  The most reductionist description of macroscopic phenomena must give a complete accounting of the quantities for current $J$, momentum $K$, charge potential $A$, and mass potential $G$ all living on a space-time $X$.  Elucidating that system of equations remains a topic of wide interest.

\begin{figure}[t]
\includegraphics[scale=.4]{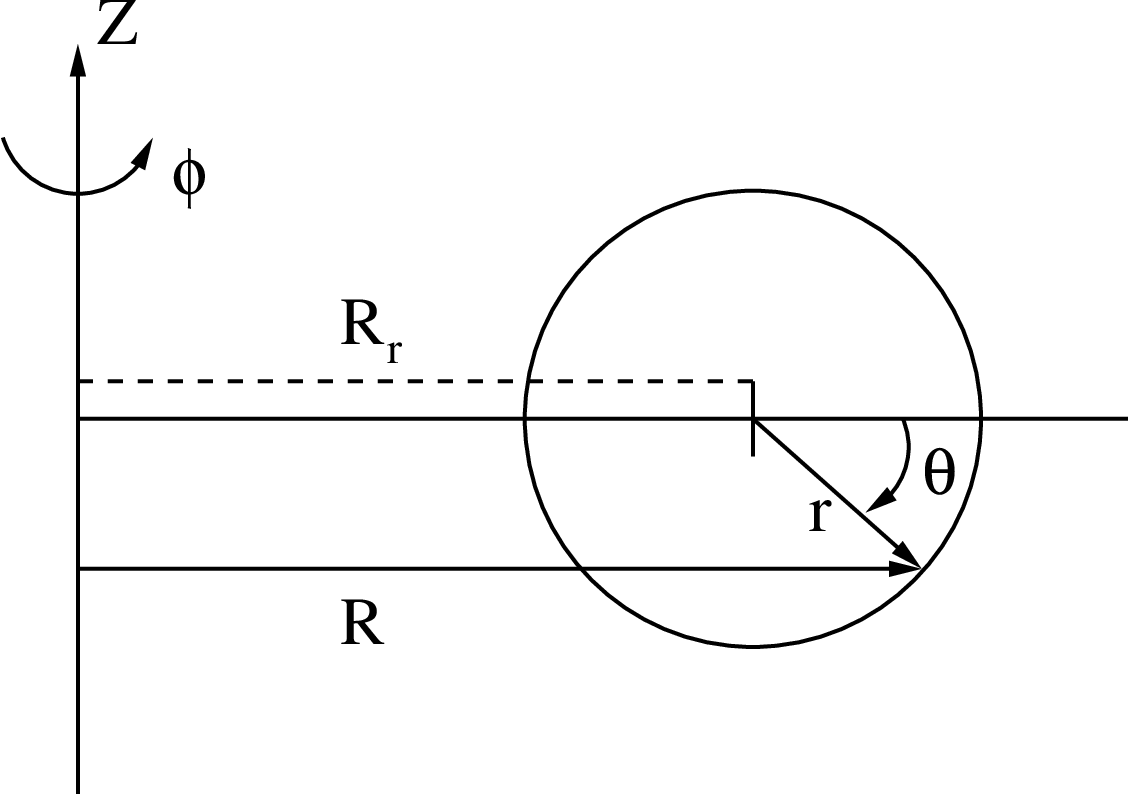}%
\caption{Tokamak coordinates $(r, \theta, \phi)$ and cylindrical coordinates $(Z, R, \phi)$. }
\label{fig:D}
\end{figure}

\appendix
\section{Useful relations} \label{app:A}
There are (at least) three useful sets of coordinate axes to describe a toroidal magnetic confinement device with concentric circular flux surfaces as shown in Figure~\ref{fig:D}, namely $(\Zhat, \Rhat, \phihat)$, $(\rhat, \thetahat, \phihat)$, and $(\rhat, \perphat, \parahat)$, and in the infinite aspect ratio limit ($R_0\rightarrow\infty$) we have the axial coordinate axes $(\Zhat, \Rhat, \zhat)$ and $(\rhat, \thetahat, \zhat)$.  Note that the term ``toroidal coordinates'' means something very different to a mathematician than those commonly applied to a tokamak, which we call ``tokamak coordinates.''  For a plasma with coaxial applied electric and magnetic fields and free current driving a circulating field, we note that $(\rhat, \thetahat, \zhat) = (-\mvec{E} \times \mvec{B}/E B, -\mvec{E} \times (\mvec{E} \times \mvec{B})/E^2 B, \mvec{E}/E)$ and $(\rhat, \perphat, \parahat) = (-\mvec{E} \times \mvec{B}/E B, -\mvec{B} \times (\mvec{E} \times \mvec{B})/E B^2, \mvec{B}/B)$.  Cylindrical coordinate labels $(Z,R, \phi)$ relate to tokamak coordinates $(r, \theta, \phi)$ via $Z=-r\Sth$ and $R=R_0+r\Cth$ in the concentric circular approximation, where $\rhat$, $\thetahat$, and $\phihat$ give the radial, poloidal, and toroidal directions, respectively.  The outermost minor radius of the confined plasma, given in meters by $a$, defines the normalized minor radius $r/a$, and $R_a$ is its centroid.

The magnetic field $\mvec{B}$ and current density $\mvec{J}$ lie in isobaric surfaces given by $\del p = \mvec{J}\times\mvec{B}$ for a stationary equilibrium, defining the ``flux surface'' at radius $r$.  In general, the nested flux surfaces are neither circular nor concentric; however, a concentric circular geometry, $R_0 = R_r$ for $r \in [0,a]$, is often used as a first approximation.  The relationship between vector components in the tokamak coordinates $(\Zhat, \Rhat, \phihat)\leftarrow(\rhat, \thetahat, \phihat)\leftarrow(\rhat, \perphat, \parahat)$ may be succinctly expressed by \beq
\left[\begin{array}{c}F_Z\\F_R\\F_\phi\end{array}\right] = \left[\begin{array}{ccc}-\Sth&-\Cth&0\\\Cth&-\Sth&0\\0&0&1\end{array}\right] \left[\begin{array}{ccc}1&0&0\\0&b_\phi&b_\theta\\0&-b_\theta&b_\phi\end{array}\right] \left[\begin{array}{c}F_r\\F_\perp\\F_\para\end{array}\right]\;, \eeq
where $\parahat \equiv \mvec{B}/B \equiv (0,b_\theta,b_\phi)$.  Various 
useful relationships are \beq
\begin{array}{lll}Z=-r\Sth\;, & \partial Z/\partial r=Z/r\;, & \partial Z/\partial\theta =-(R-R_0)\;, \\ R-R_0=r\Cth\;, & \partial R/\partial\theta=Z\;, & \partial R/\partial r=(R-R_0)/r\;, \end{array}
\eeq and for the logarithm, we have $\partial\ln R/\partial R = 1/R$ where the units on the left are carried by the differential operator and the units on the right are carried by the result, which shows that the $\ln R$ is a pure number which carries no units.

\nolinenumbers


\bibliographystyle{elsarticle-harv}

\end{document}